\documentclass[twocolumn,pra,aps,superscriptaddress,english,floatfix,longbibliography]{revtex4-1}
\usepackage{amsmath}
\usepackage{amssymb}
\usepackage{amsfonts}
\usepackage[pdftex]{graphicx}
\usepackage[dvipsnames]{xcolor}
\usepackage{subfigure} 
\usepackage[english]{babel}
\usepackage{braket}
\usepackage{color}
\usepackage{mathtools}
\usepackage{threeparttable}
\usepackage{ragged2e}
\emergencystretch=\maxdimen
\usepackage[colorlinks,linkcolor=blue,anchorcolor=blue,citecolor=blue,urlcolor=blue]{hyperref}
\usepackage{babel}
\makeatother

\begin{document}

\preprint{APS/123-QED}

\title{ Zero-change foundry compatible silicon photonics MEMS optical switch}

\author{Arkadev Roy \textsuperscript{\textdagger}}
\affiliation{Department of EECS, University of California, Berkeley, Berkeley, California, 94720}%
\thanks{These authors contributed equally to this work.}
\author{Daniel Klawson \textsuperscript{\textdagger}}
\affiliation{Department of EECS, University of California, Berkeley, Berkeley, California, 94720}%
\thanks{These authors contributed equally to this work.}

\author{Jianheng Luo}
 \affiliation{%
Department of EECS, University of California, Berkeley, Berkeley, California, 94720}%
 \author{Yiyang Zhi}
 \affiliation{%
Department of EECS, University of California, Berkeley, Berkeley, California, 94720}%
 \author{Sirui Tang}
 \affiliation{%
Department of EECS, University of California, Berkeley, Berkeley, California, 94720}%

 \author{Ming Wu}
 \email{wu@eecs.berkeley.edu}
\affiliation{%
Department of EECS, University of California, Berkeley, Berkeley, California, 94720}%

%

\begin{abstract}
Large-scale photonic switches are emerging as essential devices for energy-efficient optical interconnect in data-centers and AI/ML clusters as a key enabler for high-bandwidth and low-latency connectivity. Combining micro-electro-mechanical (MEMS) based mechanical reconfigurability with silicon photonic integrated circuits can enable large-scale, low-loss, programmable platform required for large-scale optical circuit switches. We demonstrate broadband silicon photonics MEMS switch with more than 30 dB extinction ratio operating in C-band using a zero-change foundry compatible process and Back-end-of-Line (BEOL) post-processing. The optical switch element exhibit an insertion loss of less than 1.5 dB with a low static power consumption of $\sim 20$ nW at maximum actuation voltage. Our results illustrate that MEMS-based silicon photonics modulators and phase shifters can be used alongside standard silicon photonics components seamlessly in scenarios where performances in terms of footprint, extinction ratio, broad bandwidth, and low-loss operation is of paramount importance.

\end{abstract}

\maketitle
As computing needs for AI large language models skyrocket, demanding an ever-increasing number of server clusters to perform distributed learning, there is a strong impetus to reduce latency and power consumption in datacenter communication traffic. While traditional datacenter architectures have centered around a pure packet-switching approach employing Clos topologies as the basis for large-scale networks, new topologies are increasingly emerging that incorporate optical circuit switches (OCSes) into the datacenter architecture, taking advantage of the dynamism and flexibility that electrical packet switch (EPS)–only networks lack \cite{wang2023topoopt, jouppi2023tpu}. Recently, Google announced large-scale deployment of OCSes in its datacenter networks, eliminating electrical packet-switch–based spine blocks to enable flexible topology engineering and rapid technology upgrade capability \cite{urata2022mission}. OCSes have also been integrated into AI/ML clusters, where the ability to dynamically reconfigure the topology has been shown to improve scaling, power consumption, and server utilization. Google’s OCSes are based on 3D optical MEMS technology with low optical loss and high radix \cite{jouppi2023tpu}; however, they rely on a free-space construction with discrete optical components that are assembled together \cite{urata2022mission, kim20031100}. Integrated-photonics–based optical switches promise superior switching speed, lower power consumption, and mass manufacturability, thanks to mature complementary metal-oxide-semiconductor (CMOS) technology.

Demonstrations leveraging the electro-optic \cite{qiao201732, dupuis2015design} and thermo-optic effects \cite{celo201632, ikeda2020large} exist, showing the potential of monolithic integration of optical circuit switches (OCS) based on silicon photonic integrated circuits. However, silicon’s relatively weak electro-optic effects result in modulators with a significant footprint, and thermo-optic tuning devices require high power consumption, both of which are substantial impediments for very large-scale integration. These devices often rely on cascaded 2×2 Mach-Zehnder interferometer switch units; however, maintaining low cumulative optical loss in such switch architectures has proven challenging—especially when scaling to high radices.

Microelectromechanical systems (MEMS) technology can enhance silicon photonics with building blocks that are compact, low-loss, broadband, and require very low power consumption \cite{quack2023integrated, errando2019mems, seok2016large, kim2023programmable, nagai2018silicon}. The ability to mechanically reconfigure circuits by leveraging waveguides mounted on MEMS actuators allows us to forgo the cascaded Mach-Zehnder mesh approach and instead opt for a cross-bar configuration, in which there is only a single switch element between the input and output paths. In this regard, silicon photonics–based MEMS optical switches outperform their thermo-optic and electro-optic counterparts due to the absence of cascading loss effects, which scale with the radix count. In particular, silicon photonic MEMS-based optical switches have been demonstrated to exhibit sub-microsecond switching times with substantially reduced power consumption \cite{seok2016large}, which can also be scaled to high-radix designs \cite{seok2019wafer} alongside CMOS driver integration \cite{henriksson2023large}.

The prior art in the context of silicon photonics MEMS switches, albeit CMOS compatible, has utilized customized material process stacks. To leverage the full potential of MEMS in integrated photonics, it is essential to integrate MEMS with standardized silicon photonics process technology \cite{quack2023integrated, han202132, govdeli2025integrated, takabayashi2021broadband, edinger2021silicon, edinger2023vacuum}. Here, we demonstrate a silicon photonics MEMS-based optical switch using a zero-change, commercial foundry–compatible process.

\section{Silicon Photonics MEMS actuators using BEOL metallization layers}

\begin{figure}[!h]
\centering
\includegraphics[width=0.5\textwidth]{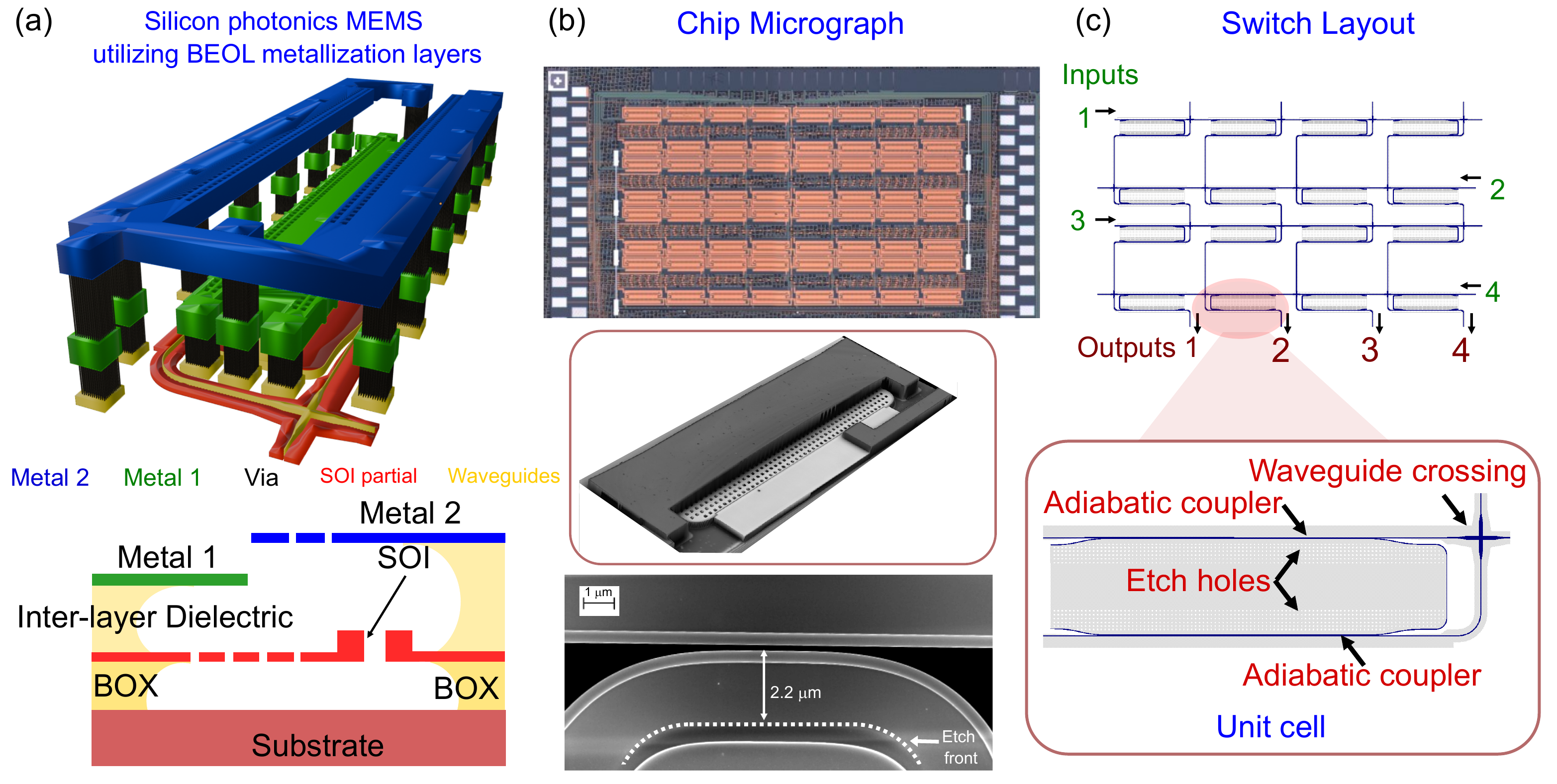}
\caption{\label{fig: schematic} \textbf{ Silicon photonic MEMS using BEOL metallization layers}. a) 3D cross-section of the silicon photonics MEMS actuator, b) Optical
micrograph of the die and zoomed-in view of a released MEMS region. Also, shown is an SEM image of a released device on-chip showing residue-free release process, c) Layout of the optical switch arranged in a modified cross-bar configuration (only a subset is shown) and the inset shows the zoomed-in view of a unit cell. }
\end{figure}

Our design was taped out through the AIM Photonics multi-project wafer (MPW) run (active PIC build). The cross-section of the MEMS design is shown in Fig. 1(a). We leverage the BEOL metallization layers to form one plate of the electrostatic MEMS actuator. The other plate of the MEMS capacitor is formed by the SOI cantilever. The nominal gap between the SOI cantilever and the Metal 2 layer is $\sim$3 $\mu$m, as set by the foundry process, and can support deflection close to 1 $\mu$m. The SOI cantilever is doped (N-doping, excluding the regions near the waveguides to eliminate impact on propagation loss), and the ground potential is applied. This acts as the global ground for all the switch elements, whereas individual addressing of each switch element is done by applying the actuation voltage to the top metal electrode. The necessary routing for supplying these voltages was done using two layers of metallization and occupies additional space between consecutive rows of the switch matrix. With the application of voltage to the MEMS actuator, the cantilever deflects upwards, transitioning from the normally ON state to the OFF configuration. The upward deflection also allows us to design for larger displacement; otherwise, a downward deflection would be limited to $\sim$650 nm, constrained by the available thickness of the buried oxide layer and the requirement to avoid pull-in. 

Figure 1(b) shows an optical micrograph of the silicon photonics die featuring an 8×8 radix optical switch. The total footprint of the radix-8 switch is 3.5 mm × 1.8 mm (with the dimensions of a unit cell measuring $\sim$ 360 $\mu$m × 130 $\mu$m). A zoomed-in view of the released MEMS region is also shown. The deflection of the MEMS is enabled by removing both the inter-metal dielectric oxide on the top as well as the buried oxide. An SEM image of a released device on the chip is shown, highlighting the residue-free release process. The switch array is arranged in a modified cross-bar configuration with low-loss waveguide crossings at waveguide intersections. The layout of the switch arranged in a cross-bar configuration is shown in Fig. 1(c). The zoomed-in view depicts the arrangement of the etch holes, which provide a pathway for the etchant to penetrate into the BOX, while the regions devoid of etch holes act as anchors defined by the timed release process (more details in Supplementary Section 2).

\section{Results}

\begin{figure}[!h]
\centering
\includegraphics[width=0.45\textwidth]{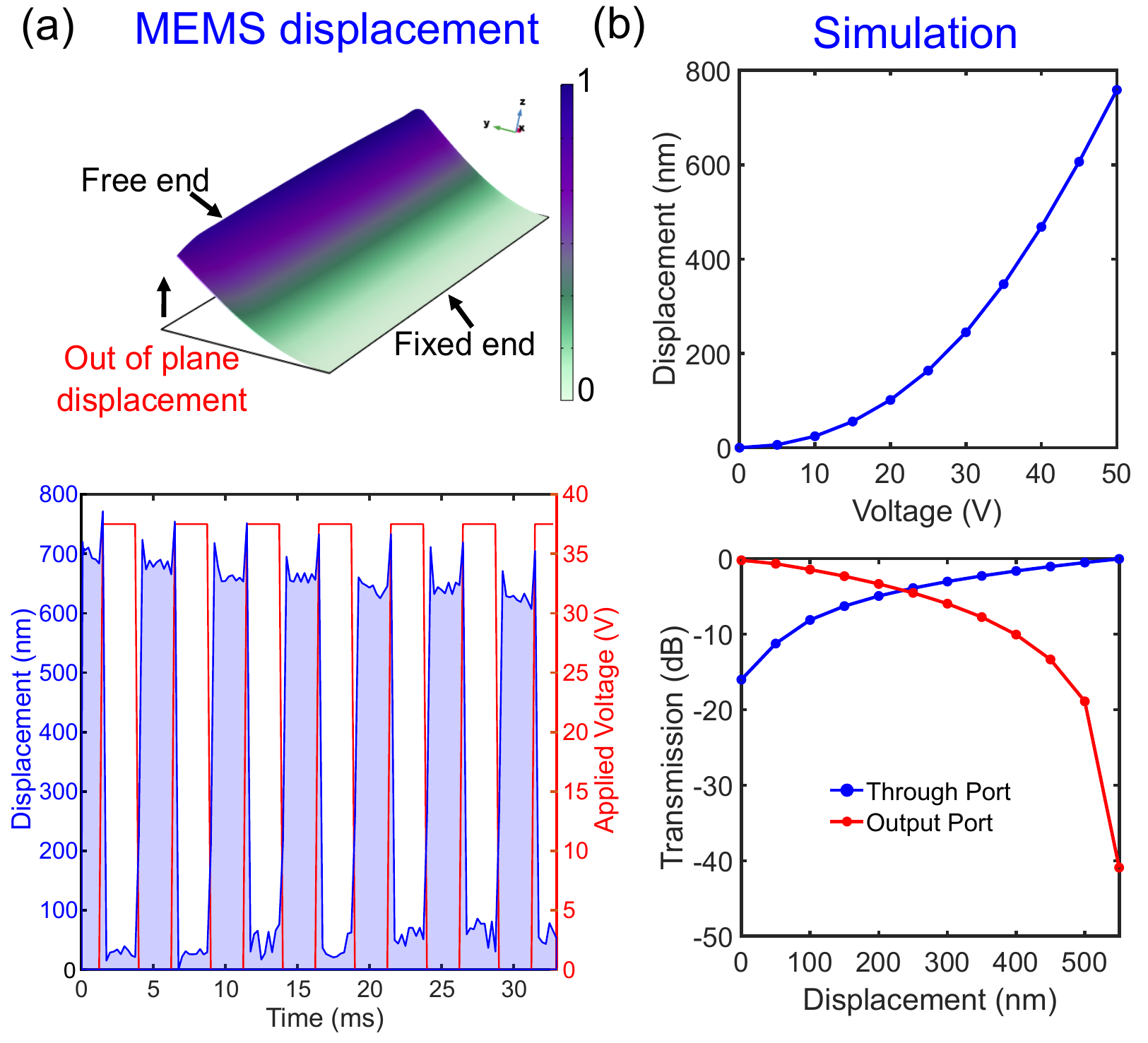}
\caption{\label{fig: schematic} \textbf{Mechanical characterization of the silicon photonics MEMS actuator}. a) MEMS displacement showing the eigenmode of the cantilever and the observed displacement through stroboscopic measurements using a Digital Holographic Microscope (DHM), b)Simulated response curves of MEMS displacement as a function of Voltage and the resulting optical transmission modulation. }
\end{figure}

The MEMS devices were released by BEOL post-processing, which involves chip-scale contact lithography followed by a timed wet release in ALPAD 639 (from Transene) and supercritical drying. The MPW die samples were coated with SPR220 resist via a spray-coating approach to a thickness of up to 9 $\mu$m. The use of thick photoresist was necessary to prevent percolation of the etchant into the masked region, which would otherwise roughen the top surface of the chip and cause excess loss in the grating couplers. Additionally, thinner resist layers were found not to survive prolonged etching, resulting in delamination from the chip midway through the process. To facilitate defect-free resist coating, we removed the top $\sim$500 nm of the passivation layer, followed by deposition of $\sim$25 nm of alumina via an atomic layer deposition process. Our choice of wet-release etch was motivated by the limited success we observed using vapor HF release, which we attribute to the presence of PECVD oxide that led to rapid undercutting of the cladding oxide (compared to the buried thermal oxide) as well as residue issues that translated to high optical loss. The buffered ALPAD etchant is selective to aluminum and copper (which are our exposed metallization layers), and the presence of surfactant facilitates surface wettability. The total etch time was estimated to be around 155 minutes (more details in Supplementary Sections 3 and 8).

The results of mechanical characterization of the MEMS actuator post-release are shown in Fig. 2. The out-of-plane displacement of the MEMS cantilever eigenmode is shown in Fig. 2(a), which is captured by stroboscopic measurements using a digital holographic microscope (bottom panel of Fig. 2(a)). We observe a deflection of $\sim$700 nm at an actuation voltage of 35 V. The simulated response curves for the MEMS actuator and its resultant effect on optical modulation are shown in Fig. 2(b). The transmission in the output port, which experiences a pair of adiabatic couplers in its path, decreases, while its through-port counterpart gradually increases and features only a single adiabatic coupler. 

\begin{figure}[!h]
\centering
\includegraphics[width=0.5\textwidth]{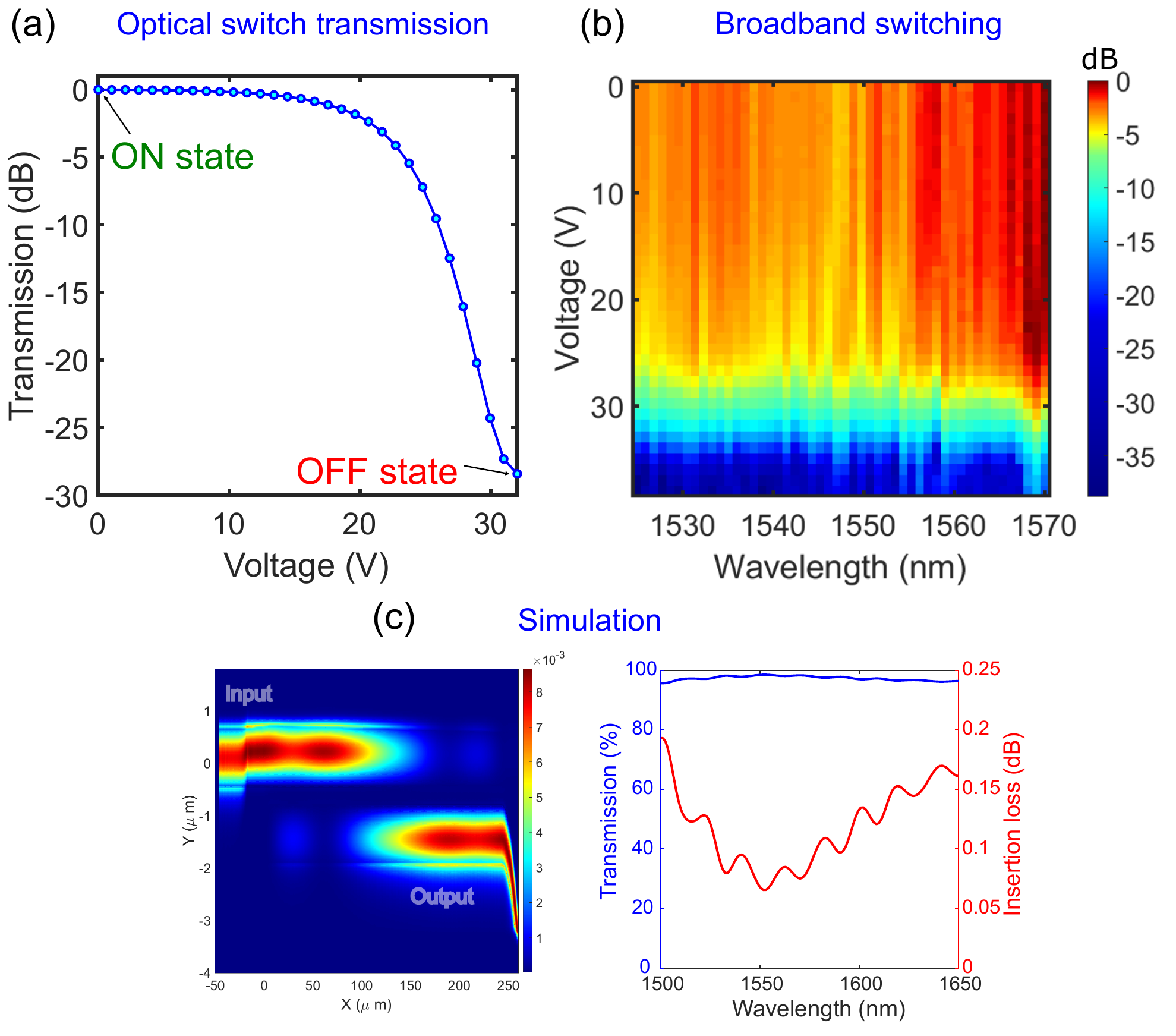}
\caption{\label{fig: schematic} \textbf{Optical characterization of the silicon photonics MEMS optical switch.} a)  Silicon photonic MEMS switch with 30 dB extinction between the ON and OFF state, b) Broadband switching capability over the C and L bands, c) Simulated field profile distribution of the adiabatic coupler (@ 1550 nm) and the transmission spectrum (insertion loss) over the bandwidth of interest.}
\end{figure}

The results of optical characterization of the silicon photonic MEMS switch unit cell are shown in Fig. 3. The active MEMS optical modulation result is shown in Fig. 3(a), characterizing the DC performance. We observe more than 30 dB extinction in the OFF state at an actuation voltage of $\sim$32 V. The broadband switching capability is shown in Fig. 3(b), which provides broadband, high-extinction-ratio switching in the C-band, supporting fat-pipe switching modality as opposed to wavelength-selective switching. The insertion loss of the switch element is measured to be less than 1.5 dB, which also includes $\sim$0.6 dB of propagation loss. The simulated performance of the designed adiabatic coupler is shown in Fig. 3(c), exhibiting an insertion loss of less than 0.25 dB over the bandwidth of interest (the switch unit cell loss includes a pair of adiabatic couplers along with contributions from waveguide propagation loss, bend loss, and four oxide-to-air transition regions). Further details of passive characterization, including waveguide crossings, are presented in Supplementary Section 5.

\begin{figure}[!h]
\centering
\includegraphics[width=0.5\textwidth]{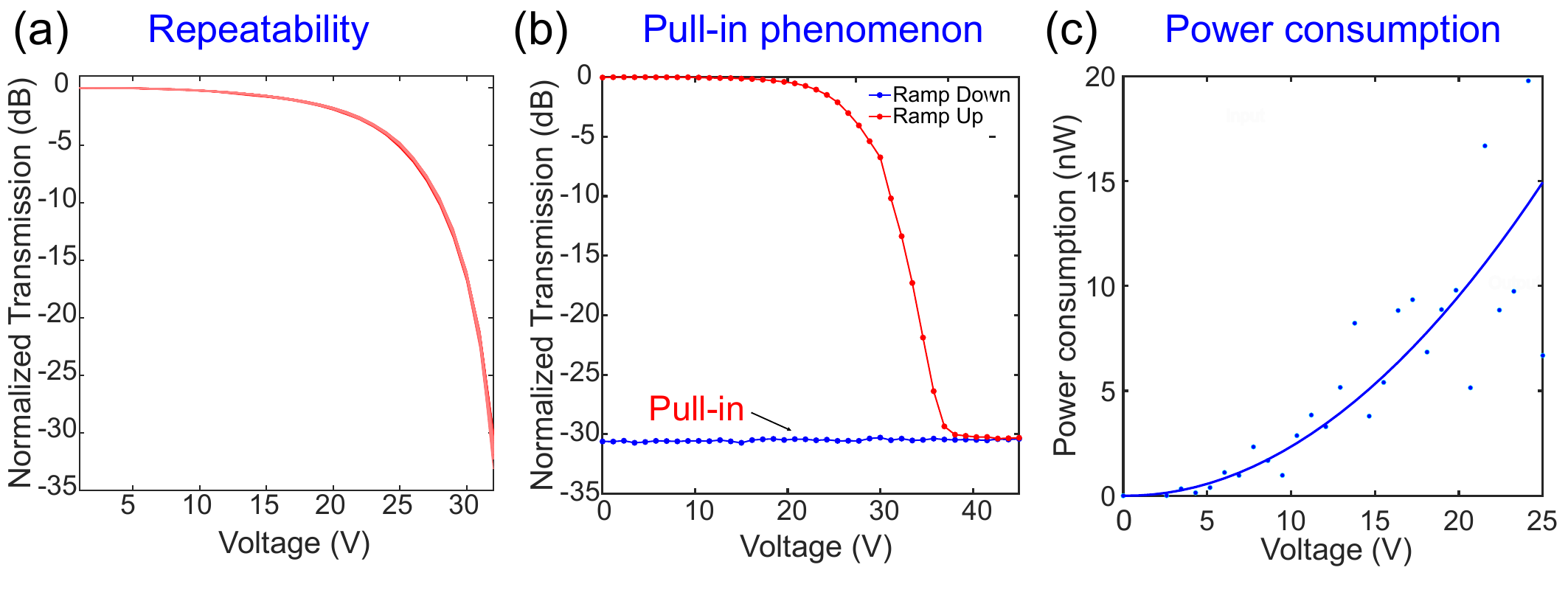}
\caption{\label{fig: schematic} \textbf{Characterizing the performance of the silicon photonics MEMS switch}. a) Repeatability of the switch transmission curves over multiple cycles depicting the analog tuning capability, b) MEMS dynamics showing the pull-in phenomenon when actuated above the pull-in threshold voltage, c) MEMS-only power consumption in steady state operation. }
\end{figure}

Further characterization of the switch unit cell is shown in Fig. 4. The repeatability of the switch transmission curves over multiple cycles (overlaid data from 100 cycles, where the switch is periodically driven and each point represents measurements taken after every 100 switching cycles) is shown in Fig. 4(a). This suggests that the device can be used for applications requiring analog tunability over a wide dynamic range, such as high-resolution analog optical matrix multiplication engines \cite{feldmann2021parallel}. Thanks to the broadband nature, it can also support massive wavelength division parallelism. The observed pull-in voltage is close to 40 V (Fig. 4(b)). Actuation above the pull-in voltage results in irreversible destruction of the MEMS actuator, which cannot be recovered. Judicious design leveraging the Metal 1 layer as a mechanical stopper can help prevent catastrophic pull-in–induced damage to some extent. The static power consumption (excluding the power consumption of the driver electronics) of the MEMS unit cell is measured to be less than 20 nW (Fig. 4(c)), arising from parasitic current dissipation through leakage pathways.

We performed temporal characterization of the optical switch to capture the switching speed (Fig. 5). We observed switching dynamics on the order of $\sim$ 600 microseconds (Fig. 5(a)). We attribute this unexpectedly slow switching time to the presence of squeezed film damping which scales with the square of the cantilever width \cite{pandey2007effect}. The damping coefficient arising from squeezed film damping can be approximated as (assuming cantilever with width $w$, length $L$, thickness $t$, and a small gap $h$ from the substrate) $C=\frac{\mu L w^{3}}{h^{3}}$, where $\mu$ is the viscosity of the fluid, which in our case is air at room temperature and atmospheric pressure. The normalized damping ratio $\zeta$ is then approximated as $\zeta=\frac{\mu W^{2}}{2\rho t \omega_{m}h^{3}}$, where $\omega_{m}$ is the resonance frequency of the cantilever. The analytically derived damping coefficient agrees well with the fit to the experimental data.

\begin{figure}[!h]
\centering
\includegraphics[width=0.5\textwidth]{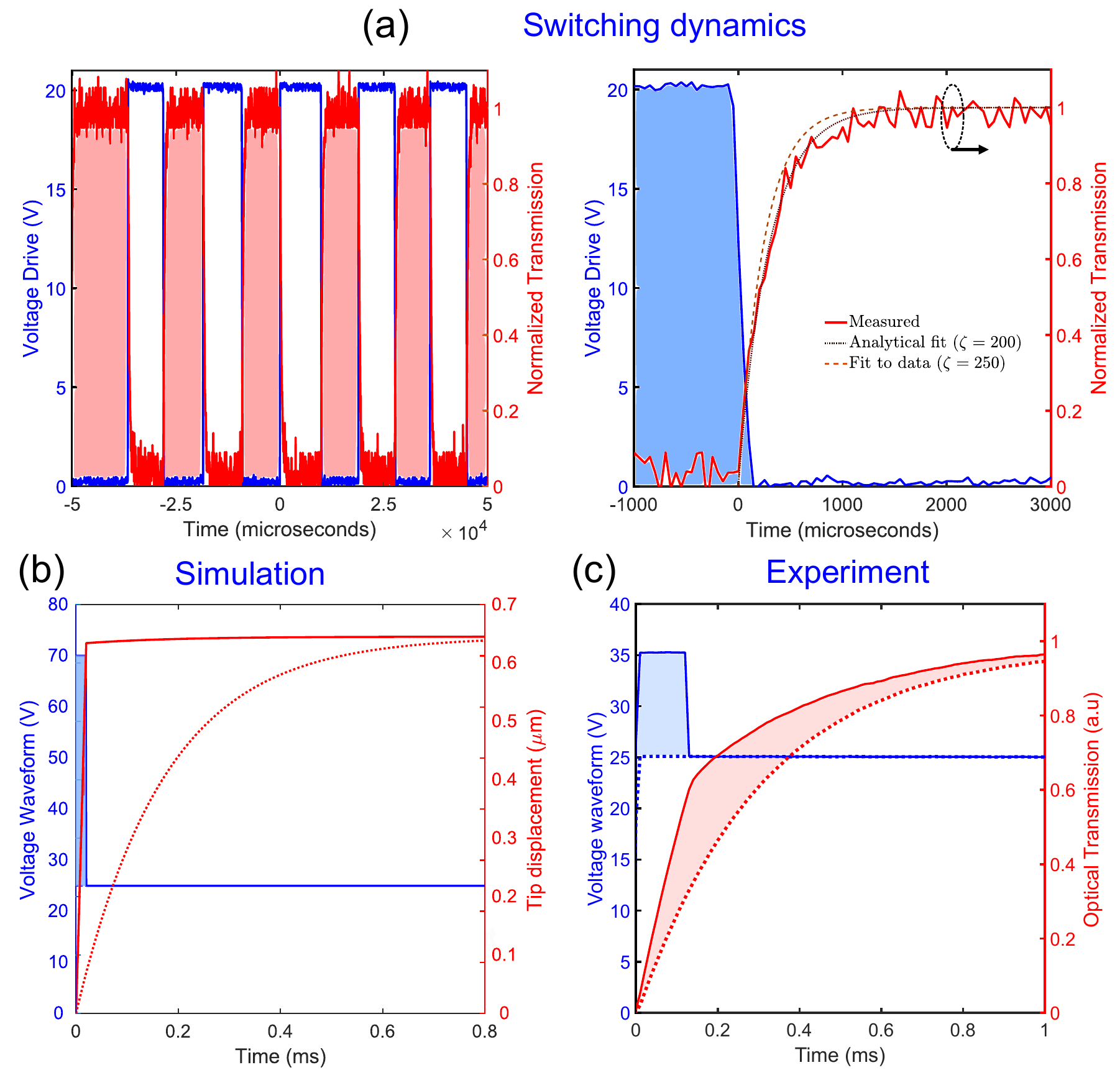}
\caption{\label{fig: schematic} \textbf{Temporal switching dynamics of the silicon photonics MEMS switch}. a) MEMS switch under a square wave excitation undergoing periodic ON to OFF transition. The zoomed-in view of the switching transient is shown alongside b) Simulated performance of the MEMS switch under dual-step high-voltage driving waveform showing accelerated switching transient compared to the single-step impulse drive, c) Measured trend of the accelerated switching dynamics under shaped high-voltage waveform driving that agrees with the simulated performance. }
\end{figure}

Regardless, the MEMS can be designed to account for this damping factor such that the MEMS is critically damped, which would support $\sim$1 microsecond of switching dynamics. The design parameters in this regard are how holey (effective surface area) the cantilever is, as well as the aspect ratio of the cantilever (width-to-length ratio), and the damping can, in principle, be tuned from the underdamped regime to the overdamped regime. Alternatively, one can resort to a hermetically sealed package with vacuum sealing to eliminate the source of squeezed film damping \cite{edinger2023vacuum}. Hermetic sealing of MEMS devices is also required for long-term reliability \cite{jo2022wafer}. The intrinsic switching dynamics are related to the mechanical resonance frequency of our SOI cantilever, which is close to 300 kHz. Nevertheless, the overdamped cantilever’s temporal response can be sped up using a shaped waveform drive that includes an overdrive voltage for a brief duration \cite{singer1990preshaping}. The shaped high-voltage waveform resembles a dual-step drive with a defined step height (i.e., the intermediate voltage level) as well as the step temporal delay, and represents an open-loop control technique. The simulated performance is shown in Fig. 5(b), highlighting the accelerated steady-state settling response compared to single-step impulse actuation (more details in Supplementary Section 6). We observe this trend in experiments (Fig. 5(c)), with a modest improvement in the temporal response, limited by the constraints of the high-speed high-voltage amplifier in our setup.

\section{Discussion}
We note that the etch rate of PECVD oxide is still higher than that of thermal oxide, even under wet etching conditions. For designs that rely on timed etching, where over-release can have deleterious effects on the anchors, we found that it is advisable to keep a larger margin for the anchor regions to prevent the MEMS structures from being washed away. The popular vapor HF-based dry release process, known for its stiction-free release and selectivity toward aluminum, can also be deployed in certain scenarios, provided that dry reactive ion-etched trenches are used to remove the majority of the cladding oxide layers down to the silicon-on-insulator layer, followed by encapsulation with a conformally atomic layer–deposited barrier material such as alumina, which is resistant to vapor HF-based release \cite{quack2023integrated, edinger2021silicon}. In our case, we could not take advantage of this approach because of the presence of metal structures that are incompatible with the foundry design rules for the deep reactive ion etching step.

We observed that the intrinsic stress in the BEOL metal layers is not prohibitive for the proper functioning of the MEMS structures. A detailed analysis of the built-in stress can be performed using stress-test structures, and the metal deposition steps can be tweaked to further optimize the MEMS actuators. Our optical switches operate in the normally ON configuration, which complicates circuit testing and requires multiple actuators to be driven simultaneously. A more practically convenient MEMS design is one in which the switch is in the normally OFF state, where the built-in metal stress can be leveraged to one’s advantage \cite{kim2023programmable}. Any residual stress in the SOI layer can be corrected by applying a global back bias to the silicon substrate. Other design choices for the MEMS actuator utilizing the BEOL metallization layers can also be explored, as discussed in Supplementary Section 1.

In summary, we have demonstrated a silicon photonic MEMS switch using a zero-change foundry process, featuring a high extinction ratio and broadband operation. The introduction of MEMS building blocks in the foundry process also opens up possibilities to integrate MEMS devices with low standby power consumption and fast switching capabilities alongside traditional active silicon photonic devices.

\begin{acknowledgments}
The authors gratefully acknowledge support from SRC JUMP 2.0 CUBIC program. The processing is done in part in the Marvell Nanolab at University of California, Berkeley and Stanford Nanofabrication Facility.
\end{acknowledgments}

\nocite{*}
\bibliography{cr}

\end{document}